\documentclass[preprint,superscriptaddress,aps]{revtex4}
\usepackage{amsfonts}
\usepackage{graphics}
\usepackage{graphicx}
\usepackage{slashed}
\newcommand{\be}{\begin{equation}}
\newcommand{\ee}{\end{equation}}
\newcommand{\en}{\end{equation}}
\newcommand{\ba}{\begin{eqnarray}}
\newcommand{\ea}{\end{eqnarray}}
\newcommand{\bea}{\begin{eqnarray}}
\newcommand{\eea}{\end{eqnarray}}

\def\pls{\partial\!\!\!/}

\def\bs{b\!\!\!/}

\def\g{\gamma}

\def\eff{\mathrm{eff}}

\begin{document}

\title{On the perturbative generation of the higher-derivative Lorentz-breaking terms}
\author{T. Mariz}
\affiliation{Instituto de F\'\i sica, Universidade Federal de Alagoas, 57072-270, Macei\'o, Alagoas, Brazil}
\email{tmariz@fis.ufal.br}
\author{J. R. Nascimento}
\affiliation{Departamento de F\'{\i}sica, Universidade Federal da Para\'{\i}ba\\
 Caixa Postal 5008, 58051-970, Jo\~ao Pessoa, Para\'{\i}ba, Brazil}
\email{jroberto, petrov@fisica.ufpb.br}
\author{A. Yu. Petrov}
\affiliation{Departamento de F\'{\i}sica, Universidade Federal da Para\'{\i}ba\\
 Caixa Postal 5008, 58051-970, Jo\~ao Pessoa, Para\'{\i}ba, Brazil}
\email{jroberto, petrov@fisica.ufpb.br}
\begin{abstract}
In this paper, we describe the perturbative generation of the higher-derivative Lorentz-breaking terms for the gauge field, that is, the Myers-Pospelov term and the higher-derivative Carroll-Field-Jackiw term. These terms are explicitly calculated in the one-loop approximation and shown to be finite and ambiguous.
\end{abstract}

\maketitle

The Lorentz symmetry breaking arises within the string context \cite{KostSam}. In the bosonic string theory the spontaneous Lorentz symmetry breaking can naturally emerge due to the instability of the naive perturbative vacuum because of the presence of the tachyon, implying that, for the correct vacuum, the vector field acquires the nontrivial vacuum expectation value introducing thus the privileged direction of the spacetime. As a result, the corresponding field theory would include the unavoidable Lorentz symmetry breaking independently of the energy scale. The similar situation can also take place in the superstrings where the role of the tachyon can be played by a dilaton \cite{KostSam}. Therefore, it is natural to suggest that the Lorentz symmetry breaking is a fundamental consequence of instability of the vacuum rather than the low-energy reduction of some fundamental Lorentz invariant theory. Moreover, following the concept of the noncommutative space-time \cite{SW}, it is natural to suggest that the fundamental theory itself must be noncommutative and therefore will not possess the Lorentz invariance. All this calls the attention to the Lorentz-breaking field theory models.

The searching for the possible Lorentz-breaking extensions of the standard model is one of the most interesting lines of study in the modern high energy physics. Within the usual approach, the Lorentz-breaking modifications are introduced as additive terms proportional to a small constant tensor introducing a privileged frame in the spacetime and thus breaking the Lorentz symmetry. A great list of the Lorentz-breaking extensions of field theory models is presented in \cite{Kostel}. The first term studied in this context is the CPT-odd Carroll-Field-Jackiw (CFJ) term \cite{CFJ}, and many issues related to it have been discussed in a number of papers \cite{list}. Another intensively studied Lorentz-breaking term is the CPT-even aether term whose perturbative generation was discussed in \cite{aether,MCN}, and some its properties at the tree level were considered in \cite{Ferr}.

From the other side, it is well known that the effective action can be represented in the form of the derivative expansion, see f.e. \cite{DE,Das}. In a general case, this expansion includes an infinite number of terms involving all possible numbers of derivatives, with the only restrictions on their structure are imposed by the gauge invariance. In the usual, Lorentz-invariant case, some higher-derivative contributions to the fermion determinant have been explicitly obtained in \cite{DE}.  Therefore, in the Lorentz-breaking extensions of the QED, the explicit calculation of the higher-derivative contributions to the one-loop effective action (that is, to the fermion determinant) is a very interesting problem. We note that, due to growing the number of possible structures with increasing the order in derivatives, the complete explicit calculation of the one-loop effective action of the gauge field seems to be impossible, thus, the only way to study it consists in computing the derivative dependent contributions order by order. In this paper, we are going to calculate the third-order contribution. Originally, the presence of higher derivatives was known to allow for essential improvement of the ultraviolet behaviour of the field theories, which plays the key role in gravity studies. Within the context of the Lorentz symmetry breaking, the first known higher-derivative term is the gravitational Chern-Simons term \cite{JaPi} which, being expressed in terms of the metric fluctuation, involves three derivatives. Its perturbative generation has been performed in \cite{ourgra}. Therefore, it is very interesting to obtain and discuss higher-derivative terms for other theories, especially, for some Lorentz-violating extensions of QED, which are expected to emerge naturally in the low-energy limit of the fundamental theories, see discussions in \cite{KostMew}. 

An important example of such a term for the vector field, called the Myers-Pospelov term, involving three derivatives, just as the gravitational Chern-Simons term, has been proposed in \cite{MP} within the phenomenological context where some estimations for the corresponding couplings were done. One of the most interesting properties clearly calling attention to this term is the fact that it generates the rotation of the polarization plane of light. The problem of causality and stability with regard to these higher-derivative terms have been discussed in \cite{Reyes}, the possible extension of the standard model by the three-derivative gauge terms, whose example is this term, has been discussed in \cite{Bolokhov}, special properties of the classical solutions in theories with such terms have been discussed in \cite{Ganguly}, and a first example of the perturbative generation based on the use of the constant third-rank tensor has been proposed in \cite{Mariz}. 

In this paper, we propose a scheme allowing for a more simple way to generate the three-derivative Lorentz-breaking terms, in particular, the Myers-Pospelov term. This scheme is based on the use of the constant vector instead of the constant third-rank tensor, in which we will see that the Myers-Pospelov term is always accompanied by a higher-derivative CFJ term.

We start with the following model which represents the spinor electrodynamics with two Lorentz-breaking couplings, that is, minimal one, proportional to $e$, and the noniminimal one, proportional to $g$ (one should notice that these couplings have different mass dimensions): 
\be
\label{mcn}
{\cal L}=\bar{\psi}\left[i \pls- \gamma^{\mu}(eA_{\mu}+g\epsilon_{\mu\nu\lambda\rho}F^{\nu\lambda}b^{\rho}) - m -  \gamma_{5}\bs\right]\psi-\frac{1}{4}F_{\mu\nu}F^{\mu\nu}.
\ee
Within this model, the $b_\rho$ is a constant vector implementing the Lorentz symmetry breaking, and $F_{\mu\nu}=\partial_{\mu}A_{\nu}-\partial_{\nu}A_{\mu}$ is the usual stress tensor corresponding to the gauge field $A_{\mu}$. The reason for choosing of this model is as follows. This model is a nonminimal extension of the well-known minimal Lorentz-breaking spinor QED characterized by the additional term $\bar{\psi}\gamma_5\bs\psi$ used in \cite{CFJ,list} to study the simplest impacts of the Lorentz breaking in gauge theories. The new term $g\bar{\psi}\epsilon_{\mu\nu\lambda\rho}F^{\nu\lambda}b^{\rho}\psi$ is CPT-odd as well as $\bar{\psi}\gamma_5\bs\psi$, therefore, the theory (\ref{mcn}) can be called the nonminimal CPT-odd Lorentz-breaking QED. We note that the possible impact from the higher-rank CPT-odd constant tensor will not essentially differ from the results of this model, see discussion in \cite{Kostel,CollMac}, therefore we restrict our discussion by the case of the inclusion of the constant vector. Since both terms have been used for the perturbative induction of the CFJ term (minimal \cite{list} and nonminimal \cite{MCN}), the above model (\ref{mcn}) is therefore the more general one for studies of the perturbative generation of the higher-derivative gauge terms. 

Besides of this, the possible extension of this model with higher-derivative terms in the fermionic sector like $\bar{\psi}\gamma_5 \slashed{b}(b\cdot\partial)^2 \psi$ and terms with more numbers of derivatives will imply in arising the divergent contributions which must be removed through the renormalization, whereas just the action (\ref{mcn}) is the only one which can yield finite contributions in the gauge sector, therefore we restrict our consideration to this action. However, we note that even in the case of this extension, both the Myers-Pospelov term and the higher-derivative CFJ term emerge. We will present these results elsewhere.

Note that in \cite{aether} the simplified version of this model involving only the nonminimal interaction was used for studies of the perturbative induction of the CPT-even aether term, and in \cite{MCN} its complete version was used for the same purpose.

The one-loop effective action of the gauge field $A_{\mu}$, denoted as  $S_\eff[b,A]$, can be expressed in the form of the following functional trace:
\be
\label{det}
S_\eff[b,A]=-i\,{\rm Tr}\,\ln(\slashed{p}-\gamma^{\mu}\tilde{A}_{\mu}- m - \gamma_5 \bs),
\ee
where 
\bea
\label{tilde}
\tilde{A}_{\mu}=eA_{\mu}+g\epsilon_{\mu\nu\lambda\rho}F^{\nu\lambda}b^{\rho}.
\eea
This effective action can be expanded in the following power series,
\be
\label{ea}
S'_\eff[b,A]=i\,{\rm Tr} \sum_{n=1}^{\infty}\frac1n
\Biggl[\frac1{\slashed{p}- m - \gamma_5 \bs}\, \g^{\mu}\tilde{A}_{\mu}\Biggr]^n.
\ee
Within our studies, we are interested only in the contributions of the second order in $\tilde{A}_{\mu}$ (further we will keep into account, among these contributions, only the terms of the third order in $b_{\mu}$). The relevant expression is given by
\be
S_\eff^{(2)}[b,A]=\frac{i}{2}{\rm Tr}\frac{1}{ \slashed{p}- m - \gamma_5 \bs}\;\g^{\mu}\tilde{A}_{\mu}\;\frac{1}{ \slashed{p}- m - \gamma_5 \bs}\;\g^{\nu}\tilde{A}_{\nu},
\ee
or, in other words,
\be
S_\eff^{(2)}[b,A]=\frac{i}{2}\int d^4x\, \Pi_b^{\mu\nu}\tilde{A}_{\mu}\tilde{A}_{\nu},
\ee
where
\be\label{Pib}
\Pi_b^{\mu\nu}={\rm tr}\int\frac{d^4p}{(2\pi)^4}\frac{1}{\slashed{p}-m-\gamma_5\slashed{b}}\gamma^\mu\frac{1}{\slashed{p}-i\slashed{\partial}-m-\gamma_5\slashed{b}}\gamma^\nu.
\ee

Using the above equations, one can find that the contribution to the one-loop effective action of the third order in $b_{\mu}$ is given by three contributions, where the number of insertions of the vector $b^{\mu}$ into the propagators is equal to one, two or three. This corresponds to two, one or zero ``nonminimal'' vertices, i.e. those ones of the form $g\bar{\psi}\epsilon_{\mu\nu\lambda\rho}\gamma^{\mu}F^{\nu\lambda}b^{\rho}\psi$, respectively. Let us consider all these situations.

First, we consider a contribution characterized by two nonminimal vertices. Using the above equation (\ref{Pib}), we must calculate the contribution given by 
\ba\label{I1}
\Pi_{b1}^{\mu\nu} &=& {\rm tr}\int\frac{d^4p}{(2\pi)^4}S(p)\gamma_5\slashed{b}S(p)\gamma^\mu S(p-k)\gamma^\nu \nonumber\\
&&+{\rm tr}\int\frac{d^4p}{(2\pi)^4}S(p)\gamma^\mu S(p-k)\gamma_5\slashed{b}S(p-k)\gamma^\nu,
\ea
with $S(p)=(\slashed{p}-m)^{-1}$, where we taken into account that $i\partial\to k$, after a Fourier transform. This expression, superficially divergent, has been calculated in many ways in the context of the perturbative generation of the CFJ term. The result obtained is $\Pi_{b1}^{\mu\nu} =C\epsilon^{\mu\nu\lambda\rho}b_\lambda k_\rho$, where the coefficient $C$ has been shown to be finite and ambiguous, depending essentially on the regularization scheme (for more details, see \cite{list}). 

In terms of this coefficient, the corresponding effective action takes the form
\be\label{term1}
S_{b1}=2g^2C\int d^4x\,  \left[b^\alpha F_{\alpha\mu}(b\cdot\partial)b_\beta\epsilon^{\beta\mu\nu\lambda}F_{\nu\lambda}+b^2b_\beta\epsilon^{\beta\mu\nu\lambda}A_\mu\Box F_{\nu\lambda}\right].
\ee
Here we have obtained the Myers-Pospelov term \cite{MP} and the higher-derivative CFJ term (first discussed in \cite{Mariz}), respectively. Note that the above terms are gauge invariant, as required for a consistent theory. 

Within our study, we specify the procedure of calculation by following the prescription of moving the $\gamma_5$ matrix to the very end of each expression which involves the trace of Dirac matrices. We note that the same result can be obtained by use of the prescription proposed by 't Hooft and Veltman \cite{Vel}. Thus, we find
\be
\Pi_{b1}^{\mu\nu} = \frac 1{2\pi^2}\left[1 - \frac{4m^2}{\sqrt{(4m^2-k^2)k^2}}\arctan\left(\frac{\sqrt{k^2}}{\sqrt{4m^2-k^2}}\right)\right]\epsilon^{\mu\nu\lambda\rho}b_\lambda k_\rho.
\ee
We consider this expression in the low-energy limit, that is, we take into account the small $k$ leading term. In this case, the $\Pi_{b1}^{\mu\nu}$ is reduced to
\be
\label{pib1}
\Pi_{b1}^{\mu\nu} = -\frac{1}{12\pi^2} \frac{k^2}{m^2}\epsilon^{\mu\nu\lambda\rho}b_\lambda k_\rho + {\cal O}\left(\frac{k^4}{m^4}\right).
\ee
Since the $\Pi_{b1}^{\mu\nu}$ is contracted with two $F_{\alpha\beta}$ which already contain space-time derivatives of the gauge field $A_{\alpha}$, we see that this contribution will correspond to the terms of fifth order in space-time, hence within this regularization the third-order terms, including the Myers-Pospelov term, do not arise from this contribution.

Now, let us consider another two contributions of third order in $b^{\mu}$, which essentially require consideration of the minimal coupling. First, one has the contribution to the effective Lagrangian involving one vertex with minimal coupling, and another vertex with nonminimal coupling. Therefore, the contribution we need to calculate from Eq.~(\ref{Pib}), is given by
\ba
\Pi_{b2}^{\mu\nu} &=& {\rm tr}\int\frac{d^4p}{(2\pi)^4}S(p)\gamma_5\slashed{b}S(p)\gamma_5\slashed{b}S(p)\gamma^\mu S(p-k)\gamma^\nu \nonumber\\
&&+{\rm tr}\int\frac{d^4p}{(2\pi)^4}S(p)\gamma_5\slashed{b}S(p)\gamma^\mu S(p-k)\gamma_5\slashed{b}S(p-k)\gamma^\nu \nonumber\\
&&+{\rm tr}\int\frac{d^4p}{(2\pi)^4}S(p)\gamma^\mu S(p-k)\gamma_5\slashed{b}S(p-k)\gamma_5\slashed{b}S(p-k)\gamma^\nu.
\ea
The calculation of the above expression is more involved because of the presence of two $\gamma_5$ matrices in the trace. The result in the naively anticommuting $\gamma_5$ scheme takes the form
\ba
\Pi_{b2}^{\mu\nu} &=& -\frac{4im^2}{\pi^2(4m^2-k^2)k^2}\left[1-\frac{(4m^2-2k^2)}{\sqrt{(4m^2-k^2)k^2}}\arctan\left(\frac{\sqrt{k^2}}{\sqrt{4m^2-k^2}}\right)\right]\nonumber \\
&&\times(b^2k^2-(b\cdot k)^2)g^{\mu\nu}.
\ea
The low-energy leading contribution of this expression is given by
\be
\Pi_{b2}^{\mu\nu} = \frac{i(b\cdot k)^2}{6\pi^2m^2}g^{\mu\nu}-\frac{ib^2k^2}{6\pi^2m^2}g^{\mu\nu}+{\cal O}\left(\frac{k^4}{m^4}\right).
\ee

After disregarding the terms of ${\cal O}(k^4/m^4)$, with the subsequent Fourier transform, we arrive at
\bea
S_{b2}=\frac{eg}{6\pi^2m^2}\int d^4x\, \left[\epsilon_{\mu\nu\lambda\rho}F^{\nu\lambda}b^{\rho}(b\cdot\partial)^2A^{\mu}-\epsilon_{\mu\nu\lambda\rho}F^{\nu\lambda}b^{\rho}b^2\Box A^{\mu}\right],
\eea
which is equivalent to
\bea
\label{term2}
S_{b2}=\frac{eg}{6\pi^2m^2}\int d^4x\, \left[b^\alpha F_{\alpha\mu}(b\cdot\partial)b_\beta\epsilon^{\beta\mu\nu\lambda}F_{\nu\lambda}+b^2b_\beta\epsilon^{\beta\mu\nu\lambda}A_\mu\Box F_{\nu\lambda}\right],
\eea
where we have taken into account that $\epsilon^{\beta\mu\nu\lambda}\partial_{\mu}F_{\nu\lambda}\equiv 0$. Note that in this case both the Myers-Pospelov term \cite{MP} and the higher-derivative CFJ term arise.

It remains to consider only the contribution to the effective Lagrangian involving both vertices with minimal coupling which requires three insertions of the $\gamma_5\bs$ into the propagator. Thus, we have
\ba
\Pi_{b3}^{\mu\nu} &=& {\rm tr}\int\frac{d^4p}{(2\pi)^4}S(p)\gamma_5\slashed{b}S(p)\gamma_5\slashed{b}S(p)\gamma_5\slashed{b}S(p)\gamma^\mu S(p-k)\gamma^\nu \nonumber\\
&&+{\rm tr}\int\frac{d^4p}{(2\pi)^4}S(p)\gamma_5\slashed{b}S(p)\gamma_5\slashed{b}S(p)\gamma^\mu S(p-k)\gamma_5\slashed{b}S(p-k)\gamma^\nu \nonumber\\
&&+{\rm tr}\int\frac{d^4p}{(2\pi)^4}S(p)\gamma_5\slashed{b}S(p)\gamma^\mu S(p-k)\gamma_5\slashed{b}S(p-k)\gamma_5\slashed{b}S(p-k)\gamma^\nu \nonumber\\
&&+{\rm tr}\int\frac{d^4p}{(2\pi)^4}S(p)\gamma^\mu S(p-k)\gamma_5\slashed{b}S(p-k)\gamma_5\slashed{b}S(p-k)\gamma_5\slashed{b}S(p-k)\gamma^\nu.
\ea
The exact form of the integral is
\ba
\Pi_{b3}^{\mu\nu} &=& \left[\frac{8m^2}{\pi^2}\frac{(2m^2-k^2)}{(4m^2-k^2)^2k^4}(b\cdot k)^2 + \frac{2}{3\pi^2}\frac{(2m^2-k^2)}{(4m^2-k^2)^2}b^2\right]\epsilon^{\mu\nu\lambda\rho}b_\lambda k_\rho \nonumber\\
&&+\left[\frac{32m^2}{3\pi^2}\frac{(6m^4+4m^2k^2-k^4)}{(4m^2-k^2)^{5/2}(k^2)^{5/2}}(b\cdot k)^2-\frac{16m^2}{3\pi^2}\frac{(m^2-k^2)}{(4m^2-k^2)^{5/2}\sqrt{k^2}}b^2\right]\nonumber\\
&&\times\arctan\left(\frac{\sqrt{k^2}}{\sqrt{4m^2-k^2}}\right)\epsilon^{\mu\nu\lambda\rho}b_\lambda k_\rho,
\ea
so that, in the low-energy limit, we arrive at
\be
\Pi_{b3}^{\mu\nu} = -\left[\frac{4(b\cdot k)^2}{45\pi^2m^4}-\frac{b^2k^2}{9\pi^2m^4}\right]\epsilon^{\mu\nu\lambda\rho}b_\lambda k_\rho+{\cal O}\left(\frac{k^4}{m^4}\right).
\ee

After the Fourier transform, we obtain
\bea
\label{term3}
S_{b3}=\frac{4e^2}{45\pi^2m^4}\int d^4x\, \left[b^\alpha F_{\alpha\mu}(b\cdot\partial)b_\beta\epsilon^{\beta\mu\nu\lambda}F_{\nu\lambda}+\frac54b^2b_\beta\epsilon^{\beta\mu\nu\lambda}A_\mu\Box F_{\nu\lambda}\right].
\eea
This term, being superficially finite, is free of any ambiguities. 

We see that the three-derivative contribution to the self-energy tensor arisen in the theory is given by the sum of the expressions (\ref{term1}), (\ref{term2}), and (\ref{term3}). Its explicit form is
\bea
\label{term}
S_{hd}&=&\left(2g^2C+\frac{eg}{6\pi^2m^2}+\frac{4e^2}{45\pi^2m^4}\right)\int d^4x\, b^\alpha F_{\alpha\mu}(b\cdot\partial)b_\beta\epsilon^{\beta\mu\nu\lambda}F_{\nu\lambda}\nonumber\\
&&+\left(2g^2C+\frac{eg}{6\pi^2m^2}+\frac{e^2}{9\pi^2m^4}\right)\int d^4x\,b^2b_\beta\epsilon^{\beta\mu\nu\lambda}A_\mu\Box F_{\nu\lambda},
\eea
with the prescription used for obtaining the Eq. (\ref{pib1}) corresponds to $C=0$. 
We see that this contribution is finite and gauge invariant. Also, we note that for the light-like Lorentz-breaking vector $b^{\mu}$, this contribution is completely described by the Myers-Pospelov term. However, we note that from the viewpoint of stability and analyticity of solutions, the space-like $b^{\mu}$ is preferable while the light-like $b^{\mu}$ leads to instabilities \cite{Reyes}. Therefore, it is natural to suggest that the higher-derivative CFJ term must arise in a consistent theory. This is also consistent with the analysis related to the minimal Lorentz-breaking spinor QED \cite{Lehnert}, since the time-like $b^{\mu}$ can produce a certain stability problem, although it is not clear whether this problem affects the consistency of the theory.

Now, it is the time to do some numerical estimations. First of all, it follows from \cite{tables} that the typical value of the components of the $b^{\mu}$ is about $10^{-32}$ GeV. Since the dimensionless constant $C$ in (\ref{term}) is of the order $\frac{1}{\pi^2}$, within all calculation schemes \cite{list}, $e\simeq 10^{-1}$ is the usual electron charge, and it is natural to suggest that $m$ is the electron mass, $m\simeq 0.5\times 10^{-3}$ GeV, as well as estimating $g\leq \frac{e}{m^2}$ to have the effect of the nonminimal interaction to be not higher that the effect of the minimal interaction, we see that we can characterize the coefficient accompanying the three-derivative term as $\frac{e^2}{m^4}b^3$, which has been parametrized in \cite{MP} by the factor $\xi/M_P$, where $M_P\simeq 10^{19}$ GeV is the Planck mass and $\xi$ is a dimensionless number. Thus, we can make the estimation $\xi/M_P\simeq 10^{-86}$ GeV$^{-1}$, i.e. $\xi\simeq 10^{-67}$. Such a small value of this number is naturally follows from the smallness of the Lorentz-breaking vector $b^{\mu}$ \cite{tables} together with the fact that the quantum correction (\ref{term}) is of the third order in this vector.

Let us briefly describe the terms with lower number of derivatives which can arise as perturbative corrections in the theory (\ref{mcn}). The analysis carried out in \cite{aether,MCN} shows that in the zero order in $b^{\mu}$ one can have the gauge-breaking Proca-like term which vanishes within an appropriate regularization prescription. In the first order in $b^{\mu}$, the CFJ term is naturally generated, and in the second order in $b^{\mu}$, one will have the aether term as it was shown in \cite{aether}.

Now, let us discuss our results. We have considered the perturbative generation of the three-derivative gauge-invariant term in the extended QED involving both minimal and nonminimal couplings. This term turns out to be gauge invariant and UV finite, reproducing a linear combination of the Myers-Pospelov term, known for the highly nontrivial manner of the propagating of solutions admitting for the rotation of plane of polarization of light \cite{MP}, and the higher-derivative CFJ term, whose contribution however vanishes for the light-like $b^{\mu}$. The ambiguity of the coefficient $C$ accompanying this term, in a pure nonminimal sector (\ref{term1}), is identically the same as that one accompanying the CFJ term in the usual Lorentz-breaking QED \cite{ourLV}. This shows that the ABJ anomaly which is known to be responsible for the ambiguity of the CFJ term \cite{JackAmb} can be naturally promoted to the higher-derivative theories. One must note, however, that this ambiguity disappears if we switch off the nonminimal interaction, therefore, it does not arise in the usual QED, although the Myers-Pospelov term arises even in this case as we have shown. 

{\bf Acknowledgements.} This work was partially supported by Conselho
Nacional de Desenvolvimento Cient\'{\i}fico e Tecnol\'{o}gico (CNPq), 
Coordena\c{c}\~{a}o de Aperfei\c{c}oamento do Pessoal
do Nivel Superior (CAPES: AUX-PE-PROCAD 579/2008) and
CNPq/PRONEX/FAPESQ. The work by A. Yu. P. has been supported by the
CNPq project No. 303461/2009-8.

\end{document}